\documentclass[aps,prb,showpacs,reprint, superscriptaddress]{revtex4-1}

\usepackage{graphicx}
\usepackage{color}
\usepackage{amsmath}
\usepackage{bbm}

\input epsf

\begin{document}

\title{Spontaneous appearance of nonzero momentum Cooper pairing: Possible application to the iron-pnictides}
\author{Micha{\l} Zegrodnik}
\email{michal.zegrodnik@agh.edu.pl}
\affiliation{Academic Centre for Materials and Nanotechnology, AGH University of Science and Technology,
Al. Mickiewicza 30, 30-059 Krakow, Poland}
\author{Jozef Spa{\l}ek}
\email{ufspalek@if.uj.edu.pl}
\affiliation{Marian Smoluchowski Institute of Physics, 
Jagiellonian University, ul. Reymonta 4,
30-059 Krak\'ow, Poland}
\affiliation{Academic Centre for Materials and Nanotechnology, AGH University of Science and Technology,
Al. Mickiewicza 30, 30-059 Krakow, Poland}

%\date{22.10.2014}

\begin{abstract}
We suggest that an inhomogeneous (non-zero momentum) paired phase can appear in the absence of an external magnetic field in the system with a predominant interband pairing and with separate Fermi-surface sheets. The Fermi wave vector mismatch which appears in such situation can be compensated by nonzero center-of-mass momentum of the Cooper pairs, what can lead to a spontaneous appearance of the Fulde-Ferrell type of superconducting state. The idea is examined using a tight-binding model which emulates the hole-like and the electron-like bands of iron based superconductor. The state can appear for the case of both spin-singlet and -triplet pairing channels.
\end{abstract}


\maketitle

\section{Introduction}
The superconducting state with non-zero center-of-mass momentum of the Cooper pairs has been proposed decades ago by Fulde and Ferrell\cite{Fulde1964} as well as independently by Larkin and Ovchinnikov\cite{Larkin1964}. According to the original idea, the so-called FFLO paired phase can be induced by an external magnetic field which gives rise to the Zeeman splitting of the Fermi surface. The Fermi wave vector mismatch between the spin subbands, is detrimental to the Cooper pair creation for electrons with opposite spins and momenta ($\mathbf{k}\uparrow$, $-\mathbf{k}\downarrow$) but may lead to an inhomogeneous paired state with a nonzero total momentum of the Cooper pairs, ($\mathbf{k}\uparrow$, $-\mathbf{k}+\mathbf{Q}\downarrow$).

It should be noted, that particular requirements are to be met to observe such supercurrent carrying state. The Maki parameter\cite{SaintJames1969} has to be large enough so that orbital effects are negligible and superconductivity survives up to the Pauli limit. Furthermore, as FFLO state is easily destroyed by the presence of impurities, the system has to be very clean. So far, there has been experimental signs of the FFLO phase in heavy fermion compounds \cite{Gloos1993}$^-$\cite{Correa2007}, as well as in organic superconductors \cite{Lee1997}$^-$ \cite{Shinagawa2007}, both belonging to the quasi-two dimensional systems. Theoretical investigations regarding the FFLO phase appearance in iron-pnictides \cite{Ptok2013_1,Ptok2014_1, Ptok2014_2} have been carried out recently as these compounds meet some of the requirements for the nonzero momentum pairing (the Pauli limit evidence in KFe$_2$As$_2$ has been reported very recently\cite{Zocco2013}).

The presence of FFLO paired state has also been proposed for imbalanced ultracold Fermi gases\cite{Zwierlein2006,Schunck2007}. The indirect experimental evidence of a superfluid FFLO phase in a system consisting of $^6$Li atoms trapped in an array of one-dimensional tubes has been reported in Ref. \onlinecite{Liao2010}. Moreover, theoretical investigations have been done concerning finite momentum Cooper pairing induced by the interplay between the artificial spin-orbit coupling and the effective Zeeman field\cite{Wu2013,Liu2013,Dong2013}. Also, the spin-triplet FFLO type of phase induced by the spin-orbit coupling has been considered for CePt$_3$Si\cite{Tanaka2007}.

Here we propose the possibility of a spontaneous (without any applied field) appearance of the FFLO type of phase. The creation of such unconventional paired state is independent of the Maki criterion for the Cooper pair decomposition. However, to make the appearance of such state possible, the interband pairing must be predominant.
Moreover, the structure of the bands has to lead to appropriate Fermi wave vector mismatch between two Fermi surface sheets. We examine this idea by using a model which consists of both the electron- and the hole-like bands of an iron-based superconductor. However, application to other inequivalent-band systems with interband pairing is also possible. If the mechanism is found operative it should be relatively easy to detect such a spontaneous supercurrent-carrying state in the superconducting-ring geometry.

It should be noted that the so-called pair density-wave state, which bears some resemblance to the Larkin-Ovchinnikov state, in zero magnetic field has been considered in Ref. \onlinecite{Loder2010} for the case of intraband pairing as well as in Ref. \onlinecite{Nikolic2010} for the case of both intra- and inter-band pairing components. However, both the origin and other features of such state are different than those considered here.

The structure of the paper is as follows. In section II we introduce the theoretical model describing electron- and hole-like bands for the iron based superconductor with proper non-zero momentum interband pairing terms. In section III we present the results of our calculations. First, we limit to the case of two hole-like bands of the considered model, to illustrate the basic features of the proposed unconventional paired phase. Then, the results for the full form of the electronic structure are discussed. Conclusions and outlook are provided in section IV.

%%%%%%%%%%%%%%%%%%%%%%%%%%%%%%%%%%%%%%%%%%%%%%%%%%%%%%%%%%%%%%%%%%%%%%%%%%%%%%%%
%%%%%%%%%%%%%%%%%%%%%%%%%%%%%%%%%%%%%%%%%%%%%%%%%%%%%%%%%%%%%%%%%%%%%%%%%%%%%%%%

\section{Model}
\subsection{Formalism}
The model is formulated for the case of the two hole-like and two electron-like bands of the iron based superconductor, LaFeAsO$_{1-x}$F$_x$. With respect to iron-pnictides, a variety of unconventional gap structures have been proposed corresponding to both the spin-singlet\cite{
Mazin2008,Kuroki2008,Chen2009,Maier2008} and the spin-triplet\cite{Maier2008,Lee2008,Dai2008} pairing. However, NMR studies \cite{Yashima2009,Jeglic2010,Michioka2010} provide an evidence for the pairing within the spin-singlet channel for this familly of compounds. According to the first principle band calculations\cite{Singh2008,Xu2008}, superconductivity in LaFeAsO$_{1-x}$F$_x$ is associated with the Fa-As layers and the Fermi surface consists of two hole pockets and two electron pockets. Moreover, the dominant contribution to the density of states near the Fermi level comes from the Fe-3$d$ orbitals. The Fe-As layer is composed of a square lattice of Fe ions with As ions in the center of each plaquette. The As ions are slightly shifted above and below the plane of the Fe lattice. The unit cell contains two Fe and two As ions. However, as it has been shown in Refs. \onlinecite{Raghu2008,Ran2009}, to describe the system one can start with the tight-binding Hamiltonian with only one Fe ion per unit cell and 
the corresponding Brillouin zone. In order to represent the original model of two Fe ions/unit cell, a folding procedure to the reduced Brillouin zone has to be carried out.

In constructing the tight-binding model we follow Raghu et al.\cite{Raghu2008}, and consider two hole-like and two electron-like bands which result from the folding procedure performed on the two hybridized bands defined by the dispersion relations
\begin{equation}
\begin{split}
E_{\mathbf{k}1}&=\frac{1}{2}\bigg(\epsilon_{\mathbf{k}1}+\epsilon_{\mathbf{k}2}-\sqrt{(\epsilon_{\mathbf{k}1}-\epsilon_{\mathbf{k}2})^2+4\epsilon^2_{\mathbf{k}12}}\bigg)\;,\\
E_{\mathbf{k}2}&=\frac{1}{2}\bigg(\epsilon_{\mathbf{k}1}+\epsilon_{\mathbf{k}2}+\sqrt{(\epsilon_{\mathbf{k}1}-\epsilon_{\mathbf{k}2})^2+4\epsilon^2_{\mathbf{k}12}}\bigg)\;,
\end{split}
 \label{eq:hyb_band}
\end{equation}
where
\begin{gather}
\nonumber
 \epsilon_{\mathbf{k}1}=-2t_1\cos k_x-2t_2\cos k_y-4t_3 \cos k_x\cos k_y\;,\\
 \nonumber
 \epsilon_{\mathbf{k}2}=-2t_2\cos k_x-2t_1\cos k_y-4t_3 \cos k_x\cos k_y\;,\\
 \nonumber
 \epsilon_{\mathbf{k}12}=-4t_4 \sin k_x \sin k_y\;.
 \end{gather}
 The values of the hopping parameters in the units of $|t_1|$ are as follows: $t_1=-1$, $t_2=1.3$, and $t_3=t_4=-0.85$. The effect of the folding procedure on the starting Hamiltonian which consists of the hybridized bands $E_{\mathbf{k}l}$ can be written in the following manner
 \begin{equation}
  \sum_{\mathbf{k}l\sigma} E_{\mathbf{k}l}\hat{n}_{\mathbf{k}l\sigma}\rightarrow \sideset{}{'}\sum_{\mathbf{k}l'\sigma}\tilde{E}_{\mathbf{k}l'}\hat{n}_{\mathbf{k}l'\sigma}\;,  
  \label{eq:folding}
 \end{equation}
where $l=1,2$ correspond to the two hybridized bands, whereas $l'$ correspond to the two hole-like ($l'=1,2$) and the two electron-like ($l'=3,4$) bands. The summation on the left-hand side of (\ref{eq:folding}) is over the unfolded Brillouin zone, which corresponds to one Fe ion per unit cell, while the primed summation on the right-hand side is over the folded Brillouin zone which is twice smaller and corresponds to two Fe ions per unit cell (for detailed description of the folding procedure see Refs. \onlinecite{Lee2008} and \onlinecite{Raghu2008}).
 The band structure and the Fermi surfaces are shown for the sake of clarity in Fig. \ref{fig:1}. One should notice that the Fermi-wave-vector mismatch between the two hole-like bands seen in Fig. \ref{fig:1}, is analogous to that appearing between the spin-up and the spin-down Fermi surface sheets in the presence of magnetic field for the case of one-band system with the spin-singlet pairing. Such situation represents a canonical case for introducing the FFLO state \cite{Fulde1964,Larkin1964}. The picture is slightly different for the case of the two electron-like bands as the mismatch between the Fermi sheets varies significantly with the direction in the reciprocal space. 

Next, we add to the model the term responsible for the creation of the interband Cooper pairs, i.e.,
 \begin{equation}
\begin{split}
\mathcal{\hat{H}}=\sideset{}{'}\sum_{\mathbf{k}l'\sigma}(\tilde{E}_{\mathbf{k}l'}&-\mu)\hat{n}_{\mathbf{k}l'\sigma}-\frac{2}{N}\sideset{}{'}\sum_{\mathbf{k}\mathbf{k}'mm'\mathbf{Q}}U^{mm'}_{\mathbf{k}-\mathbf{k}'}\hat{B}^{\dagger}_{\mathbf{k}'m'\mathbf{Q}}\hat{B}_{\mathbf{k}m\mathbf{Q}}\;,\\
\end{split}
  \label{eq:H_start}
 \end{equation}
where $U^{mm'}_{\mathbf{k}-\mathbf{k}'}$ is the pairing strength, $N$ is the number of Fe atoms in the layer, $\mathbf{Q}$ are the Cooper pair momenta, while $\hat{B}^{\dagger}_{\mathbf{k}m\mathbf{Q}}$ are the spin-singlet pairing operators 

\begin{equation}
\left\{\begin{array}{cc}
\hat{B}^{\dagger}_{\mathbf{k},1\mathbf{Q}}&\equiv\frac{1}{\sqrt{2}}(\hat{c}^{\dagger}_{\mathbf{k}1\uparrow}\hat{c}^{\dagger}_{-\mathbf{k}+\mathbf{Q}2\downarrow
} -\hat{c}^ { \dagger
}_{\mathbf{k}1\downarrow}\hat{c}^{\dagger}_{-\mathbf{k}+\mathbf{Q}2\uparrow})\;,\\
\\
\hat{B}^{\dagger}_{\mathbf{k},2\mathbf{Q}}&\equiv\frac{1}{\sqrt{2}}(\hat{c}^{\dagger}_{\mathbf{k}3\uparrow}\hat{c}^{\dagger}_{-\mathbf{k}+\mathbf{Q}4\downarrow
} -\hat{c}^ { \dagger
}_{\mathbf{k}3\downarrow}\hat{c}^{\dagger}_{-\mathbf{k}+\mathbf{Q}4\uparrow})\;.
\end{array}\right.
\label{eq: A_op}
\end{equation}
They correspond to pairing between the two hole-like bands (m=1) and two electron-like bands (m=2).
Similar term as the one seen in (\ref{eq:H_start}) has been introduced by Dai et al. \cite{Dai2008} but for the spin-triplet interband pairing. Also, the model considered in Ref. \onlinecite{Dai2008} was limited to the case of two electron-like bands of LaFeAsO$_{1-x}$F$_x$, and the possibility of nonzero $\mathbf{Q}$ vector has not been considered there. 

Analogously as in the original work by Fulde and Ferrell\cite{Fulde1964}, we assume that all Cooper pairs have the same momentum $\mathbf{Q}$ which leads to the gap parameters phase oscillations in real space (the FF phase). To simplify the situation, we assume that $U^{11}_{\mathbf{k}}=U^{22}_{\mathbf{k}}\equiv U_{\mathbf{k}}$ and $U^{12}_{\mathbf{k}}=U^{21}_{\mathbf{k}}\equiv \gamma U_{\mathbf{k}}$, where $\gamma$ is a constant parameter. For $\gamma=0$ the pairing within the hole-like bands would be separated from the one in the electron-like bands which would lead to the appearance of two critical temperatures. As such situation is not observed in experiment, it is necessary to 
introduce the 
pairing term 
which corresponds to the case of $m\neq m'$ ($\gamma\neq 0$), and results in a single critical temperature in the system. The term for $m\neq m'$ represents the transfer of the Cooper pairs from the electron-like to the hole-like bands and vice versa.

 As in Ref. \onlinecite{Dai2008}, we use the following form of the pairing strength $U_{\mathbf{k}}$ 
\begin{equation}
U_{\mathbf{k}}=U_0+U_1(\cos k_x+\cos k_y)\;.
\label{eq:pair_strngth} 
\end{equation}
By using the mean field (BCS) approximation one obtains the following form of the effective Hamiltonian
\begin{equation}
 \begin{split}
\mathcal{\hat{H}}_{HF}&=\sideset{}{'}\sum_{\mathbf{k}l'\sigma}(\tilde{E}_{\mathbf{k}l'}-\mu)\hat{n}_{\mathbf{k}l'\sigma}\\
&+\sqrt{2}\sideset{}{'}\sum_{\mathbf{k},m}\big[(\Delta_{\mathbf{k}m\mathbf{Q}}+\gamma\Delta_{\mathbf{k}\bar{m}\mathbf{Q}})\hat{B}^{\dagger}_{\mathbf{k}m\mathbf{Q}}+H.C.\big]\\
&+\frac{N}{U_0}((\Delta^{(0)}_{1\mathbf{Q}})^2+2\gamma\Delta^{(0)}_{1\mathbf{Q}}\Delta^{(0)}_{2\mathbf{Q}}+(\Delta^{(0)}_{2\mathbf{Q}})^2)\\
&+\frac{2N}{U_1}((\Delta^{(1)}_{1\mathbf{Q}})^2+2\gamma\Delta^{(1)}_{1\mathbf{Q}}\Delta^{(1)}_{2\mathbf{Q}}+(\Delta^{(1)}_{2\mathbf{Q}})^2)\;,
 \end{split}
\label{eq:H_HF}
\end{equation}
where $\bar{m}=2$ for $m=1$ and vice versa. In the above expression we have introduced the interband superconducting gap parameters for the two hole-like bands (m=1) and two electron-like bands (m=2)
\begin{equation}
\Delta_{\mathbf{k}m\mathbf{Q}}=-\frac{1}{\sqrt{2}}\frac{2}{N}\sideset{}{'}\sum_{\mathbf{k}'}U_{\mathbf{k}-\mathbf{k}'}\langle\hat{B}_{\mathbf{k}'m\mathbf{Q}}\rangle\;.
\end{equation}
The gap has a mixture of s-wave and extended s-wave symmetries, i.e.,
\begin{equation}
\Delta_{\mathbf{k}m\mathbf{Q}}=\Delta^{(0)}_{m\mathbf{Q}}+\Delta^{(1)}_{m\mathbf{Q}}(\cos k_x + \cos k_y)\;.
\end{equation}
The extended s-wave gap symmetry for the case of spin-singlet pairing has been also considered in Ref. \onlinecite{Mazin2008,Kuroki2008}, whereas for the case of interband spin-triplet pairing in Ref. \onlinecite{Dai2008}.

Diagonalization of (\ref{eq:H_HF}) leads to the quasiparticle energies 
\begin{equation}
\begin{split}
  \lambda_{1,2}=&\frac{1}{2}\bigg[(\tilde{E}_{\mathbf{k}1}-\tilde{E}_{-\mathbf{k}+\mathbf{Q}2})\\
  &\pm\sqrt{(\tilde{E}_{\mathbf{k}1}+\tilde{E}_{-\mathbf{k}+\mathbf{Q}2}-2\mu)^2+4(\Delta_{\mathbf{k}1\mathbf{Q}}+\gamma\Delta_{\mathbf{k}2\mathbf{Q}})^2}\bigg],\\
    \lambda_{3,4}=&\frac{1}{2}\bigg[(\tilde{E}_{\mathbf{k}3}-\tilde{E}_{-\mathbf{k}+\mathbf{Q}4})\\
  &\pm\sqrt{(\tilde{E}_{\mathbf{k}3}+\tilde{E}_{-\mathbf{k}+\mathbf{Q}4}-2\mu)^2+4(\Delta_{\mathbf{k}2\mathbf{Q}}+\gamma\Delta_{\mathbf{k}1\mathbf{Q}})^2}\bigg],\\
\end{split}
\label{eq:quasipart_ene}
\end{equation}
which are used to derive the free-energy functional in a standard statistical-mechanical manner. The gap amplitudes $\Delta_{m\mathbf{Q}}^{(0)}$ and $\Delta_{m\mathbf{Q}}^{(1)}$, and the chemical potential are all obtained by solving the set of self-consistent equations numerically, while the vector $\mathbf{Q}$ is determined by minimizing the free energy of the system.

\subsection{Methodological remarks}
The essential point of the whole analysis is the observation that the dominant interband pairing may by the only reasonable cause for the Fermi wave vector mismatch to occur spontaneously. So, even though we carry out the discussion on example of the electronic structure appropriate for the pnictides, the principal point may have a more general meaning. 

Note that the considered here pairing is of spin-singlet, band-triplet type. Within this model, spin-triplet band-singlet pairing may also be considered. In such scenario the Hund's rule can play the role of the pairing mechanism \cite{Spalek2001,Zegrodnik2012,Zegrodnik2013,Spalek2013,Zegrodnik2014}. Both of these situations represent even-parity pairing and in fact, by applying the pairing term of the spin-triplet type
 \begin{equation}
\begin{split}
\hat{H}_p=-\frac{2}{N}\sideset{}{'}\sum_{\mathbf{k}\mathbf{k}'mm'\mathbf{Q}}J^{mm'}_{\mathbf{k}-\mathbf{k}'}\hat{A}^{\dagger}_{\mathbf{k}'m'\mathbf{Q}}\hat{A}_{\mathbf{k}m\mathbf{Q}}\;,\\
\end{split}
  \label{eq:H_start_tr}
 \end{equation}
 with 
 \begin{equation}
\left\{\begin{array}{cc}
\hat{A}^{\dagger}_{\mathbf{k},1\mathbf{Q}}&\equiv\frac{1}{\sqrt{2}}(\hat{c}^{\dagger}_{\mathbf{k}1\uparrow}\hat{c}^{\dagger}_{-\mathbf{k}+\mathbf{Q}2\downarrow
} +\hat{c}^ { \dagger
}_{\mathbf{k}1\downarrow}\hat{c}^{\dagger}_{-\mathbf{k}+\mathbf{Q}2\uparrow})\;,\\
\\
\hat{A}^{\dagger}_{\mathbf{k},2\mathbf{Q}}&\equiv\frac{1}{\sqrt{2}}(\hat{c}^{\dagger}_{\mathbf{k}3\uparrow}\hat{c}^{\dagger}_{-\mathbf{k}+\mathbf{Q}4\downarrow
} +\hat{c}^ { \dagger
}_{\mathbf{k}3\downarrow}\hat{c}^{\dagger}_{-\mathbf{k}+\mathbf{Q}4\uparrow})\;,
\end{array}\right.
\label{eq: A_op}
\end{equation}
one obtains identical set of self-consistent equations (after using the mean field BCS approximation), as well as the expression for the free energy. This is caused by the fact that within our approach, in the absence of magnetic field, the spin and band indices are interchangeable. So the spin-triplet band-singlet scenario is equivalent to the spin-singlet band-triplet one. The situation is obviously different in the case with the intraband pairing included when the choice of the symmetry of the spin-part of the Cooper pairs wave function determines the symmetry of the superconducting gap in $\mathbf{k}$ space. That is why the intra-band, spin-triplet pairing is considered always in connection with odd-parity pairing (p-wave, f-wave). Here, because the Cooper pairs are formed by electrons from different bands the spin-triplet even-parity pairing (s-wave, d-wave) is also permissible.

%%%%%%%%%%%%%%%%%%%%%%%%%%%%%%%%%%%%%%%%%%%%%%%%%%%%%%%%%%%%%%%%%%%%%%%%%%%%%%%%
%%%%%%%%%%%%%%%%%%%%%%%%%%%%%%%%%%%%%%%%%%%%%%%%%%%%%%%%%%%%%%%%%%%%%%%%%%%%%%%%

\section{Results and Discussion}
For modeling purposes, we set $U_1=U_0/5$, the energies are all normalized to the bare band-width $W$, $T$ corresponds to the reduced temperature $T\equiv k_BT/W$, and the wave vectors are given in the units of $1/a$, where $a$ is the lattice parameter. The following phases are considered: (i) normal state ({\bf NS}): $\Delta^{(0)}_{m\mathbf{Q}}=\Delta^{(1)}_{m\mathbf{Q}}\equiv0$, (ii) homogeneous superconducting state ({\bf SC}): $\Delta^{(0)}_{m\mathbf{Q}}\neq 0$, $\Delta^{(1)}_{m\mathbf{Q}}\neq0$, $\mathbf{Q}=(0,0)$, and (iii) inhomogeneous superconducting state ({\bf FF}): $\Delta^{(0)}_{m\mathbf{Q}}\neq 0$, $\Delta^{(1)}_{m\mathbf{Q}}\neq 0$,  $\mathbf{Q}\neq(0,0)$. In the subsequent discussion $n$ expresses the number of electrons per one Fe ion. To investigate the general features of the proposed unconventional phase, as well as to illustrate the mechanism of its appearance, we initially limit solely to the two-hole like bands of the model described in the previous section. Next, the results for the 
full form of Hamiltonian (\ref{eq:H_start}  ) will be described.

%%%%%%%%%%%%%%%%%%%%%%%%%%%%%%%%%%%%%%%%%%%%FIG1%%%%%%%%%%%%%%%%%%%%%%%%%%%%%%%%
\begin{figure}[h!]
%\hfill
\centering
\epsfxsize=90mm 
{\epsfbox[96 444 579 661]{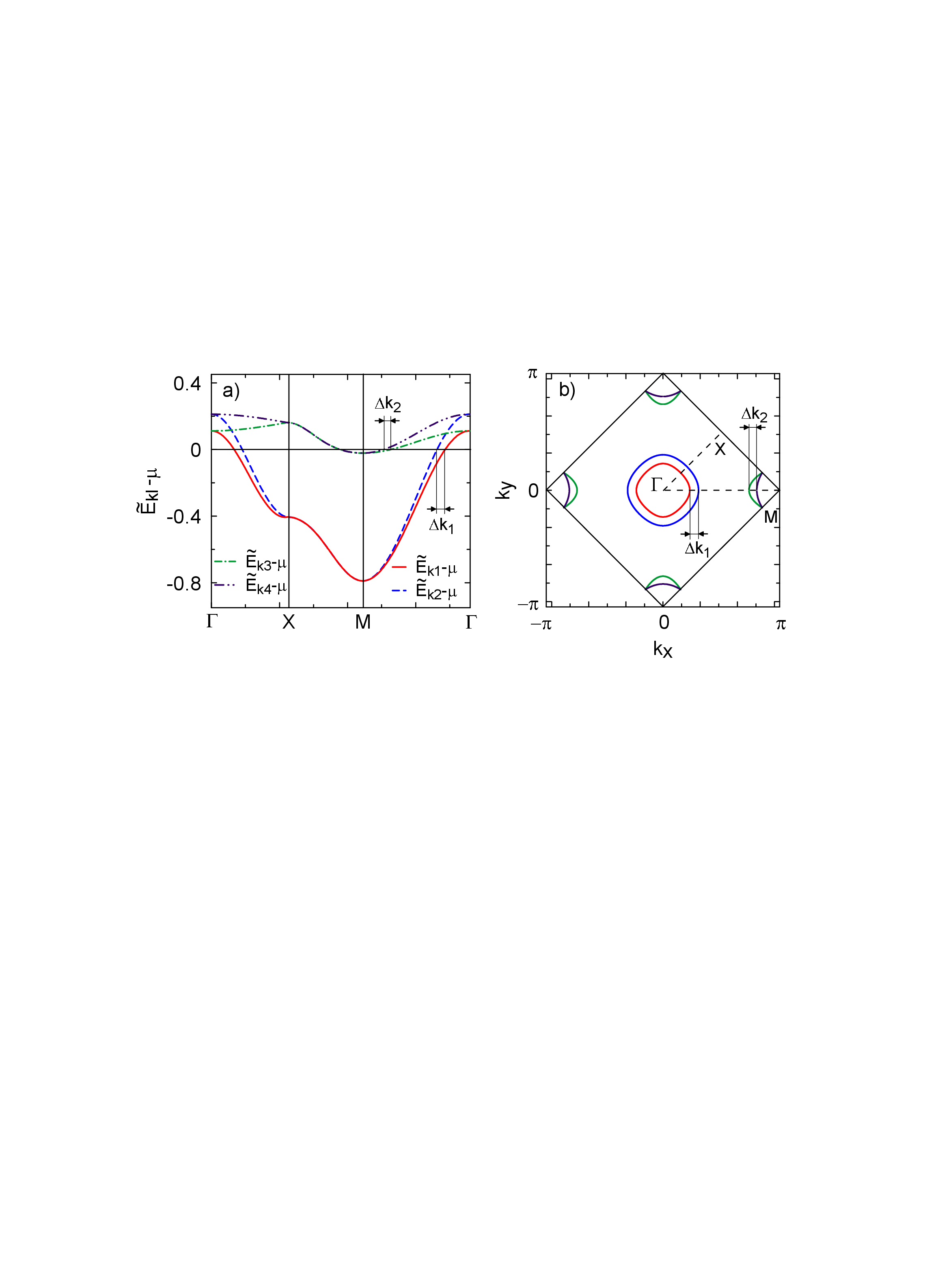}}
\caption{Electronic structure (a) and the Fermi surface sheets (b) for $n=1.94$. The black solid line in (b) marks the boundaries of the reduced Brillouin zone. The $\Delta k_1\approx0.236$ and $\Delta k_2\approx 0.210$ parameters represent the Fermi wave-vector mismatch between the hole- and electron-like bands, respectively.}
\label{fig:1}
\end{figure}
%%%%%%%%%%%%%%%%%%%%%%%%%%%%%%%%%%%%%%%%%%%%%%%%%%%%%%%%%%%%%%%%%%%%%%%%%%%%%%%%
%%%%%%%%%%%%%%%%%%%%%%%%%%%%%%%%%%%%%%%%%%%%FIG2%%%%%%%%%%%%%%%%%%%%%%%%%%%%%%%%
\begin{figure}[h!]
%\hfill
\centering
\epsfxsize=90mm 
{\epsfbox[76 527 558 685]{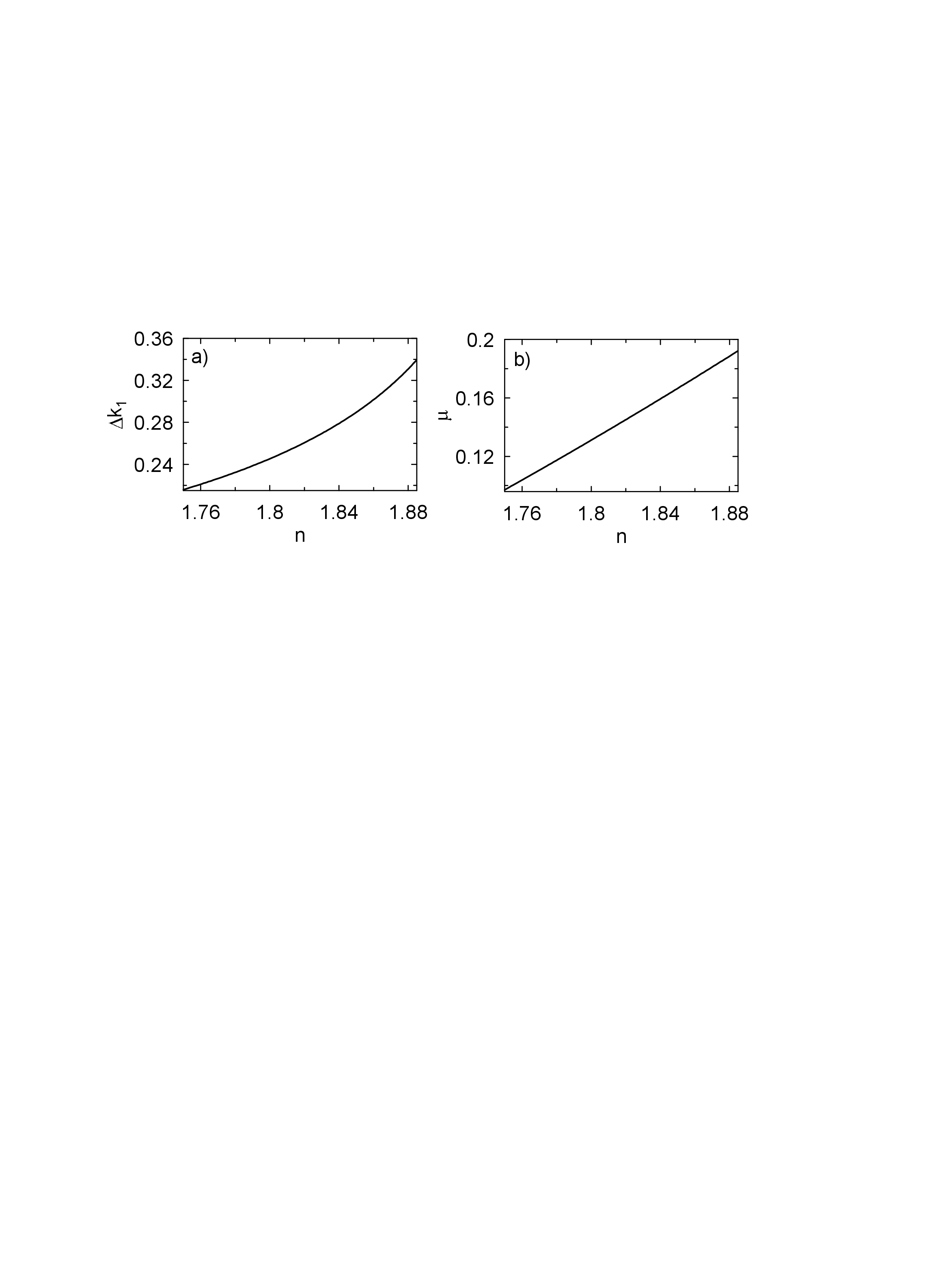}}
\caption{The $\Delta k_1$ parameter (a) and the chemical potential (b), both vs. the total band filling $n$ for the limited case of two hole-like bands in the model.}
\label{fig:2}
\end{figure}
%%%%%%%%%%%%%%%%%%%%%%%%%%%%%%%%%%%%%%%%%%%%%%%%%%%%%%%%%%%%%%%%%%%%%%%%%%%%%%%%
\subsection{Two hole-like-band case}
For the limited case of the two hole-like bands the nonzero momentum of the Cooper pairs, formed by electrons from different bands, results from the Fermi wave-vector mismatch $\Delta k_1$ between the two Fermi surface sheets centered at the $\Gamma$ point (Fig. \ref{fig:1}b). Doping the system with electrons increases the Fermi-energy, which in turn increases $\Delta k_1$, as shown in Fig. \ref{fig:2}a. Hence, by doping one can tune the $\Delta k_1$ parameter and as a result, trigger the FF phase onset in a similar manner as the conventional intraband FF-phase is triggered by the applied magnetic field. In the latter case a shift between the spin-up and spin-down dispersion relations appears. So the bottom of the bands between which the pairing occurs 
corresponds to different energies. In the present case of interband pairing the Fermi wave-vector mismatch results from a specific electronic structure. The reference points of the bands, between which the pairing occurs coincide, but different shapes of the two dispersion relations lead to the two distinct Fermi surface sheets (cf. Fig. \ref{fig:1}). In principle, a similar nonzero momentum phase, of the spin-singlet type (or the spin-triplet type, depending on the system at hand and specified pairing mechanism), could appear without magnetic field for the case of pairing between any two types of particles with different (effective) masses. For example, in the case of heavy fermion systems the Fermi wave vector mismatch is induced by both the Zeeman splitting and the different effective masses. For those compounds in the external magnetic field the spin-dependence of the quasiparticle masses appears as a result of the correlations (the mass renormalization factors are different for spin-up and spin-down 
particles). The last 
mechanism leads to the enlargement of the stability region of the FFLO phase 
on the $(T,H_a)$ plane\cite{Kaczmarczyk2010,Maska2010}. It would also be interesting the investigate if the scenario considered by us would be possible in the case of pairing between two species of particles (with two different masses) within the ultracold Fermi gas trapped in an optical lattice.
%%%%%%%%%%%%%%%%%%%%%%%%%%%%%%%%%%%%%%%%%%%%FIG3%%%%%%%%%%%%%%%%%%%%%%%%%%%%%%%%
\begin{figure}[t!]
%\hfill
\centering
\epsfxsize=85mm 
{\epsfbox[35 312 566 763]{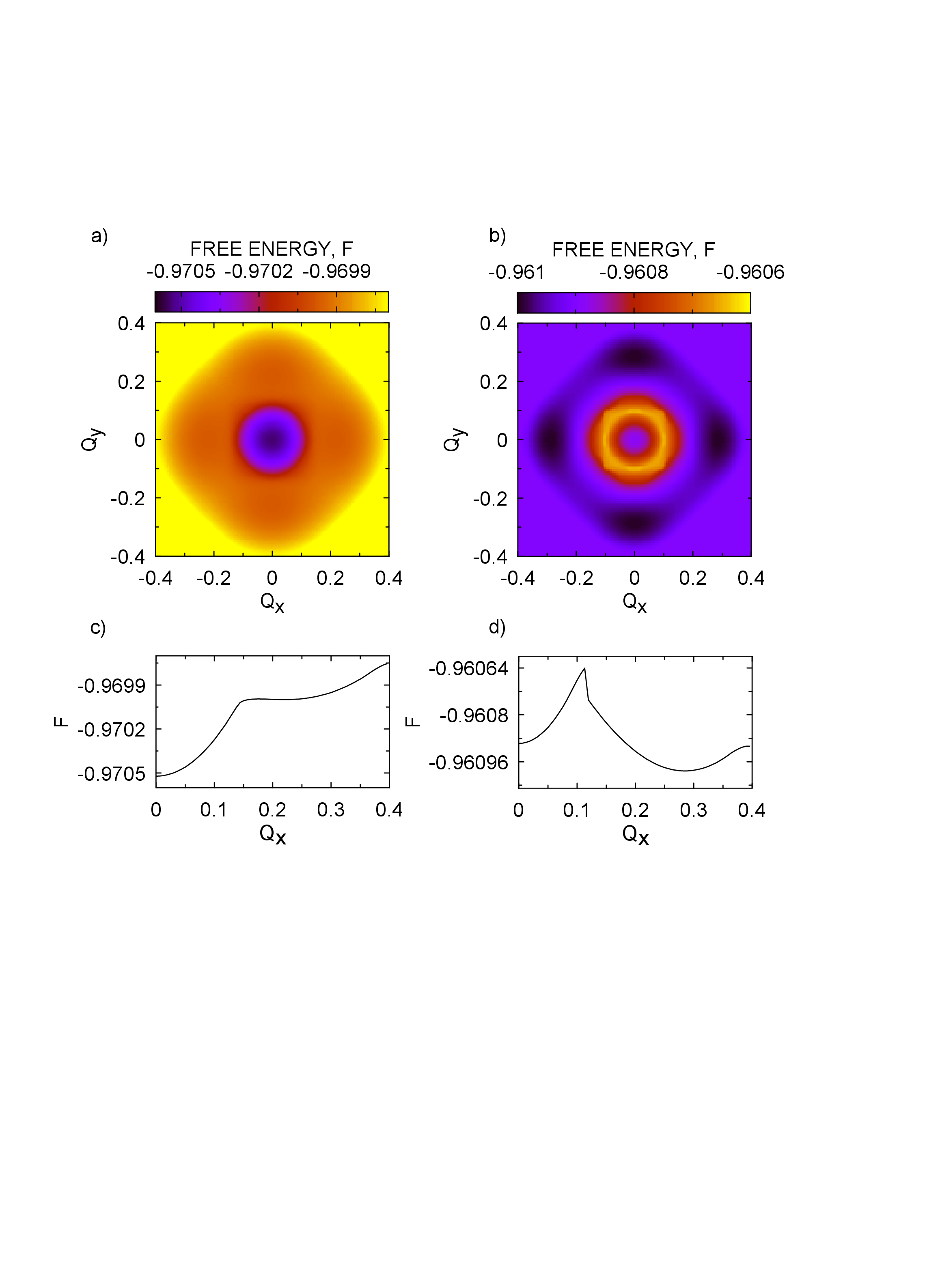}}
\caption{Free energy of the paired state as a function of the Cooper pair total momentum for $n=1.75$ (a) and $n=1.79$ (b), for the limited case of two hole-like bands only.  The pairing strength is set to $U_0=0.39$. Note the four distinct free-energy minima in (b) which signal the FF phase appearance.}
\label{fig:3}
\end{figure}
%%%%%%%%%%%%%%%%%%%%%%%%%%%%%%%%%%%%%%%%%%%%%%%%%%%%%%%%%%%%%%%%%%%%%%%%%%%%%%%%
%%%%%%%%%%%%%%%%%%%%%%%%%%%%%%%%%%%%%%%%%%%%FIG4%%%%%%%%%%%%%%%%%%%%%%%%%%%%%%%%
\begin{figure}[h!]
%\hfill
\centering
\epsfxsize=75mm 
{\epsfbox[218 530 462 664]{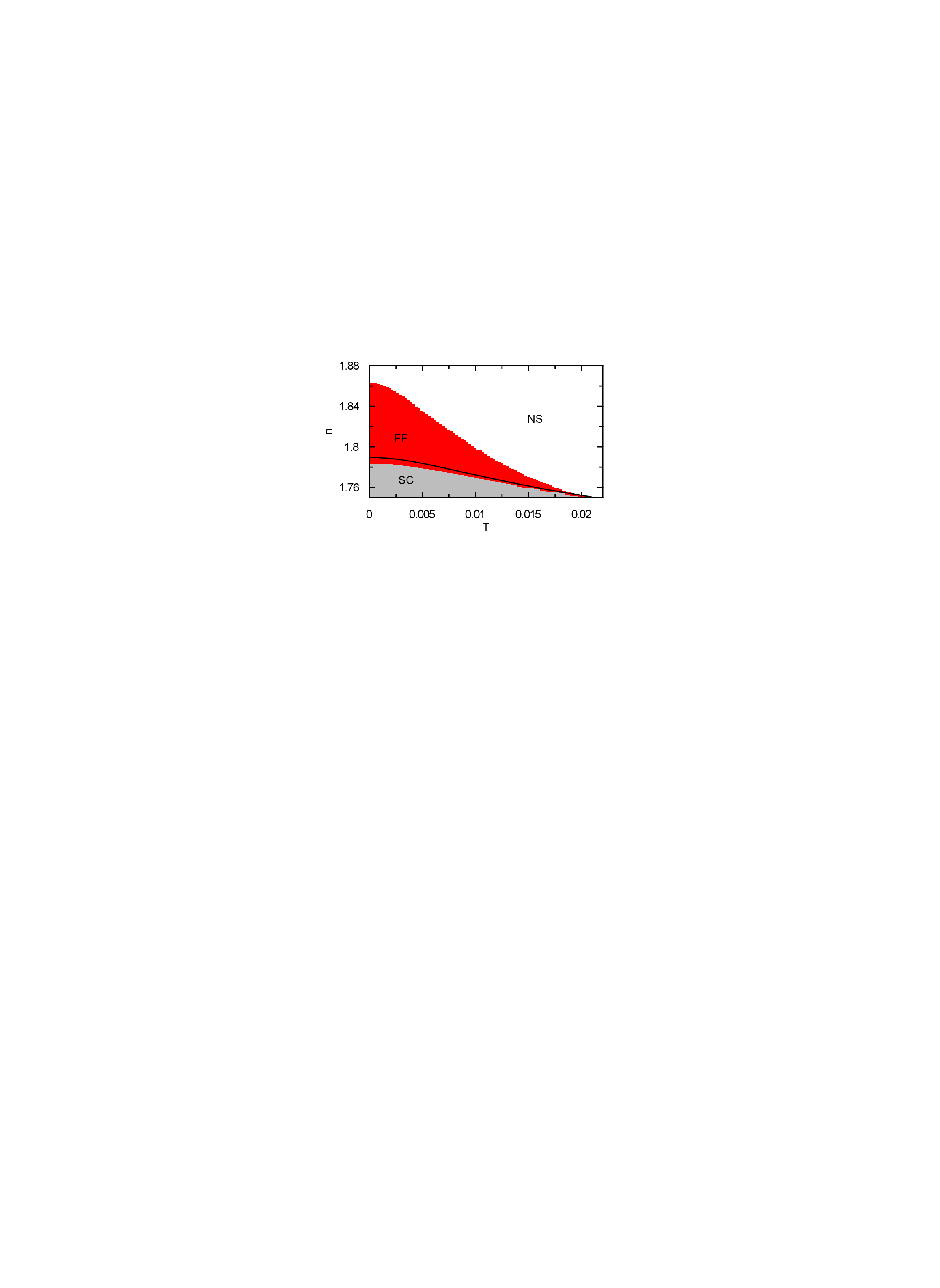}}
\caption{Phase diagram on $(T, n)$ plane for $U_0=0.39$ for the limited case of two hole-like bands only. Labels of the phases are explained in main text. The transition from  homogeneous (SC) to inhomogeneous (FF) spin-singlet phase is discontinuous. The solid line marks the border between the SC and NS phases in the case of absence of the FF phase.}
\label{fig:4}
\end{figure}
%%%%%%%%%%%%%%%%%%%%%%%%%%%%%%%%%%%%%%%%%%%%%%%%%%%%%%%%%%%%%%%%%%%%%%%%%%%%%%%%

In Fig. \ref{fig:3} we plot the free-energy for the superconducting phase as a function of the total momentum of the Cooper pairs, $\mathbf{Q}$, for two values of the band filling specified. It can be seen that for $n=1.75$ the $\Delta k_1$ parameter is too small to trigger the creation of stable FF phase, as there is only one minimum for $\mathbf{Q}=(0,0)$. However, upon increasing slightly the band filling to $n=1.79$, the free-energy minima corresponding to four distinct $\mathbf{Q}$ vectors appear, which refer to a stable FF phase. The positions of those minima are determined by the value of the mismatch, $\Delta k_1\approx0.237$ in this case, and by the shape if the Fermi surface. It is most convenient for the $\mathbf{Q}$ 
vector to 
be parallel to either the $k_x$ or the $k_y$ axis. In such situation the largest parts of the two concentric hole-like Fermi sheets can be connected. As a result, one can assume $Q_y=0$ and determine only $Q_x$ by minimizing the free energy. It should be noted that due to the cubic anisotropy and the chosen form of the pairing interaction given by Eq. (\ref{eq:pair_strngth}) one should expect that the free energy minima would appear along the high symmetry directions. However, for the sake of completeness and to check the correctness of the numerical procedure we have made calculations determining the full $(Q_x,Q_y)$ dependence which is shown in Figs. \ref{fig:3}a and \ref{fig:3}b.
%%%%%%%%%%%%%%%%%%%%%%%%%%%%%%%%%%%%%%%%%%%%FIG5%%%%%%%%%%%%%%%%%%%%%%%%%%%%%%%%
\begin{figure}[h!]
%\hfill
\centering
\epsfxsize=67mm 
{\epsfbox[131 202 477 621]{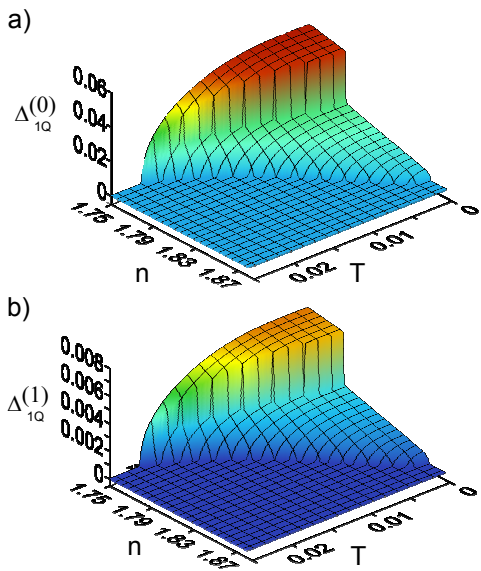}}
\caption{Gap amplitudes $\Delta^{(0)}_{1\mathbf{Q}}$ and $\Delta^{(1)}_{1\mathbf{Q}}$ as a function of both $T$ and $n$, for $U_0=0.39$ for the limited case of two hole-like bands only. Note the discontinuity in the gap parameter at the SC$\rightarrow$FF transition, while the FF$\rightarrow$NS transition is continuous.}
\label{fig:5}
\end{figure}
%%%%%%%%%%%%%%%%%%%%%%%%%%%%%%%%%%%%%%%%%%%%%%%%%%%%%%%%%%%%%%%%%%%%%%%%%%%%%%%%

\begin{figure}[h!]
%\hfill
\centering
\epsfxsize=65mm 
{\epsfbox[131 389 443 608]{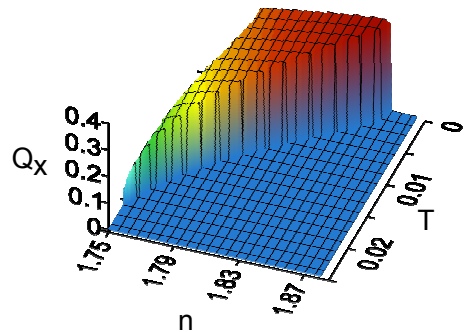}}
\caption{The center-of-mass momentum component $Q_x$ of the Cooper pairs (for $Q_y=0$) which corresponds to stable FF phase as a function of both the reduced temperature and the band filling with the parameter $U_0=0.39$. These results correspond to the limited case of two hole-like bands only}
\label{fig:6}
\end{figure}

\begin{figure}[h!]
%\hfill
\centering
\epsfxsize=75mm 
{\epsfbox[248 371 479 501]{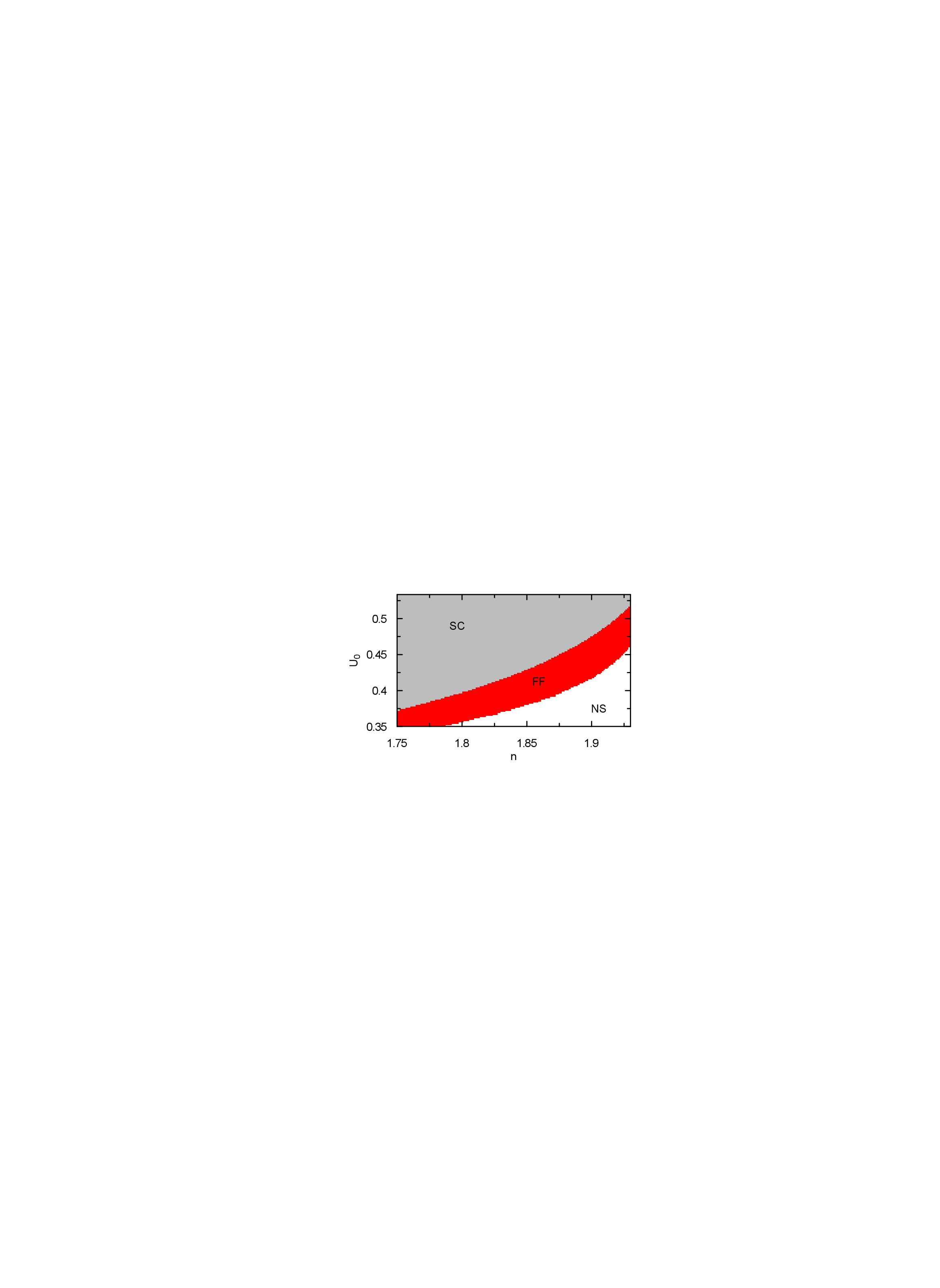}}
\caption{The phase diagram on $(n,U_0)$ plane for $U_1=U_0/5$ and for the limited case of two hole-like bands. Labels representing the phases are explained in main text.}
\label{fig:7}
\end{figure}

The phase diagram on $(T,n)$ plane is shown in Fig. \ref{fig:4}, where the regions of stability of the paired phases both homogeneous (SC) and inhomogeneous (FF) are specified. It should be noted that the band filling range, for which the FF phase is stable, corresponds to the chemical potential values which lead to Fermi surface sheets of similar size and shape, as those presented in Refs. \cite{Xu2008,Singh2008,Raghu2008}. The transition from the SC phase to FF phase has a discontinuous nature as can be seen in Fig. \ref{fig:5}, where the drops in both $\Delta^{(0)}_{1\mathbf{Q}}$ and $\Delta^{(1)}_{1\mathbf{Q}}$ are clearly visible. The $\Delta^{(1)}_{1\mathbf{Q}}$ amplitude is about one order of magnitude smaller than $\Delta^{(0)}_{1\mathbf{Q}}$. However, the critical temperature is common for both, as it should be. The $Q_x$ component of the total momentum which minimizes the free-energy and leads to stability of the FF phase is displayed in Fig. \ref{fig:6} vs. $T$ and $n$. For the sake of 
completeness, we show in Fig. \ref{fig:7} the phase 
diagram 
on $(n,U_0)
$ plane, which 
demonstrates the stable phase evolution with increasing $U_0$ and $U_1$. As one can see, the larger the band filling is, the stronger the coupling has to be to stabilize the FF paired phase. This is caused by the fact that with increasing $n$ the $\Delta k_1$ parameter also increases. 

%%%%%%%%%%%%%%%%%%%%%%%%%%%%%%%%%%%%%%%%%%%%FIG6%%%%%%%%%%%%%%%%%%%%%%%%%%%%%%%%
%%%%%%%%%%%%%%%%%%%%%%%%%%%%%%%%%%%%%%%%%%%%%%%%%%%%%%%%%%%%%%%%%%%%%%%%%%%%%%%%
%%%%%%%%%%%%%%%%%%%%%%%%%%%%%%%%%%%%%%%%%%%%FIG7%%%%%%%%%%%%%%%%%%%%%%%%%%%%%%%%
%%%%%%%%%%%%%%%%%%%%%%%%%%%%%%%%%%%%%%%%%%%%%%%%%%%%%%%%%%%%%%%%%%%%%%%%%%%%%%%%

\subsection{Full four-band case}

Here, we discuss the results for the full 4-band Hamiltonian given by Eq. (\ref{eq:H_start}). All the results presented here have been obtained with the $\gamma$ parameter set to $0.01$. In such model, one could in general consider two different $\mathbf{Q}$ vectors. One adjusted to compensate the Fermi wave vector mismatch between the hole-like bands and the other to compensate the mismatch between the electron-like bands. However, as a single critical temperature has to appear in the model, the term which corresponds to the Cooper-pair transfer from the electron- to the hole-like bands, and vice versa, needs to be introduced ($\gamma\neq 0$). 
Two independent $\mathbf{Q}$ vectors corresponding to the two types of bands appearing simultaneously would lead to interband processes for which the center-of-mass momentum of the Cooper 
pairs is not conserved. It is assumed that such processes do not appear, so the Cooper pairs can have only one value of the total momentum. 
%%%%%%%%%%%%%%%%%%%%%%%%%%%%%%%%%%%%%%%%%%%%FIG8%%%%%%%%%%%%%%%%%%%%%%%%%%%%%%%%
\begin{figure}[h!]
%\hfill
\centering
\epsfxsize=85mm 
{\epsfbox[35 312 566 763]{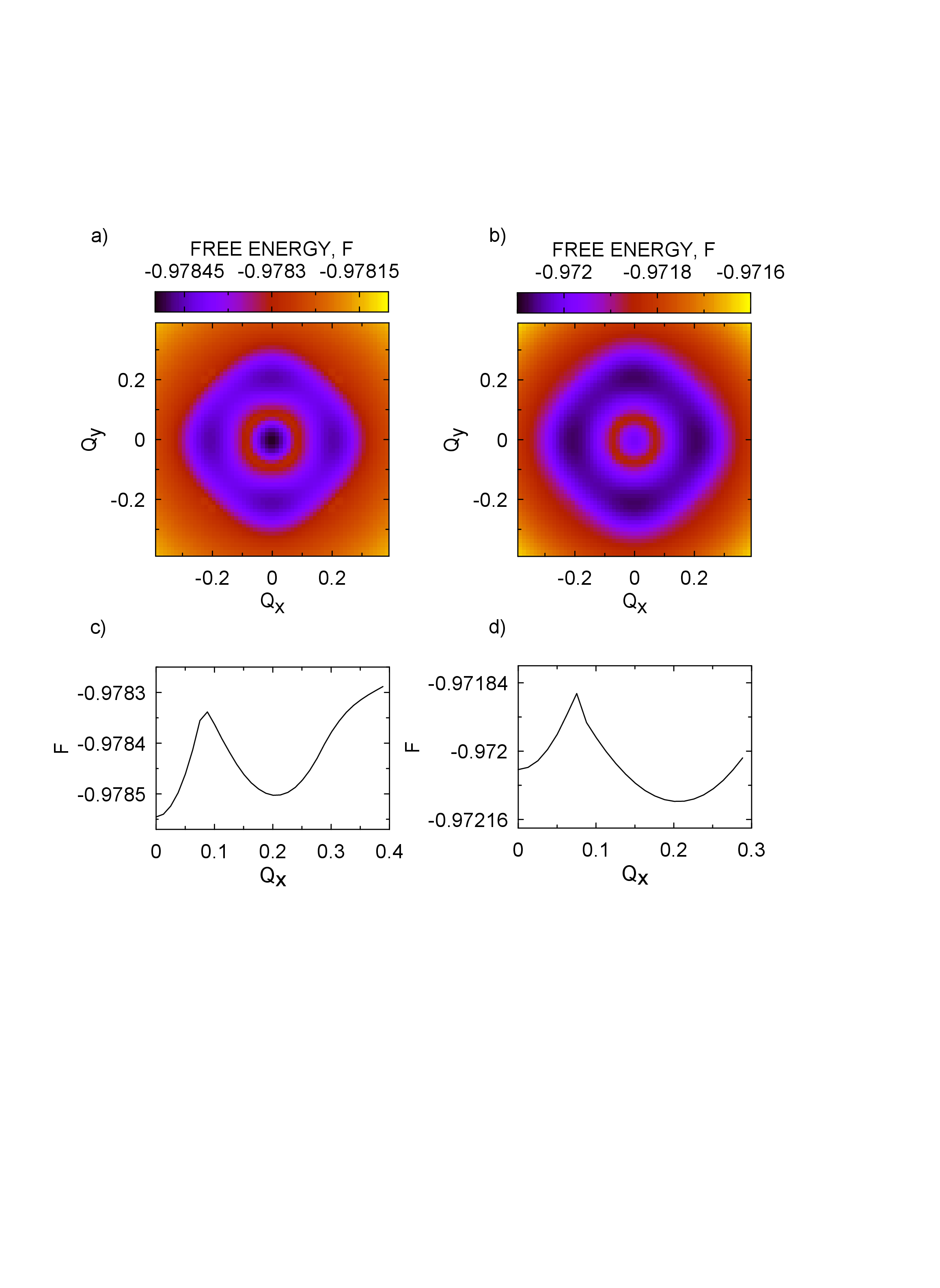}}
\caption{Free energy of the system in the paired state as a function of the Cooper pair total momentum for $n=1.72$ (a) and $n=1.795$ (b) for the model with two electron- and two hole-like bands. The pairing strength is set to $U_0=0.33$. In (c) and (d) we show the $Q_x$ dependence of the free energy for $Q_y=0$. As one can see from (b), for a proper value of band filling, the minima of the free energy appear for non-zero values of the Cooper pair momentum.}
\label{fig:8}
\end{figure}
%%%%%%%%%%%%%%%%%%%%%%%%%%%%%%%%%%%%%%%%%%%%%%%%%%%%%%%%%%%%%%%%%%%%%%%%%%%%%%%%

\begin{figure}[h!]
%\hfill
\centering
\epsfxsize=75mm 
{\epsfbox[145 249 537 629]{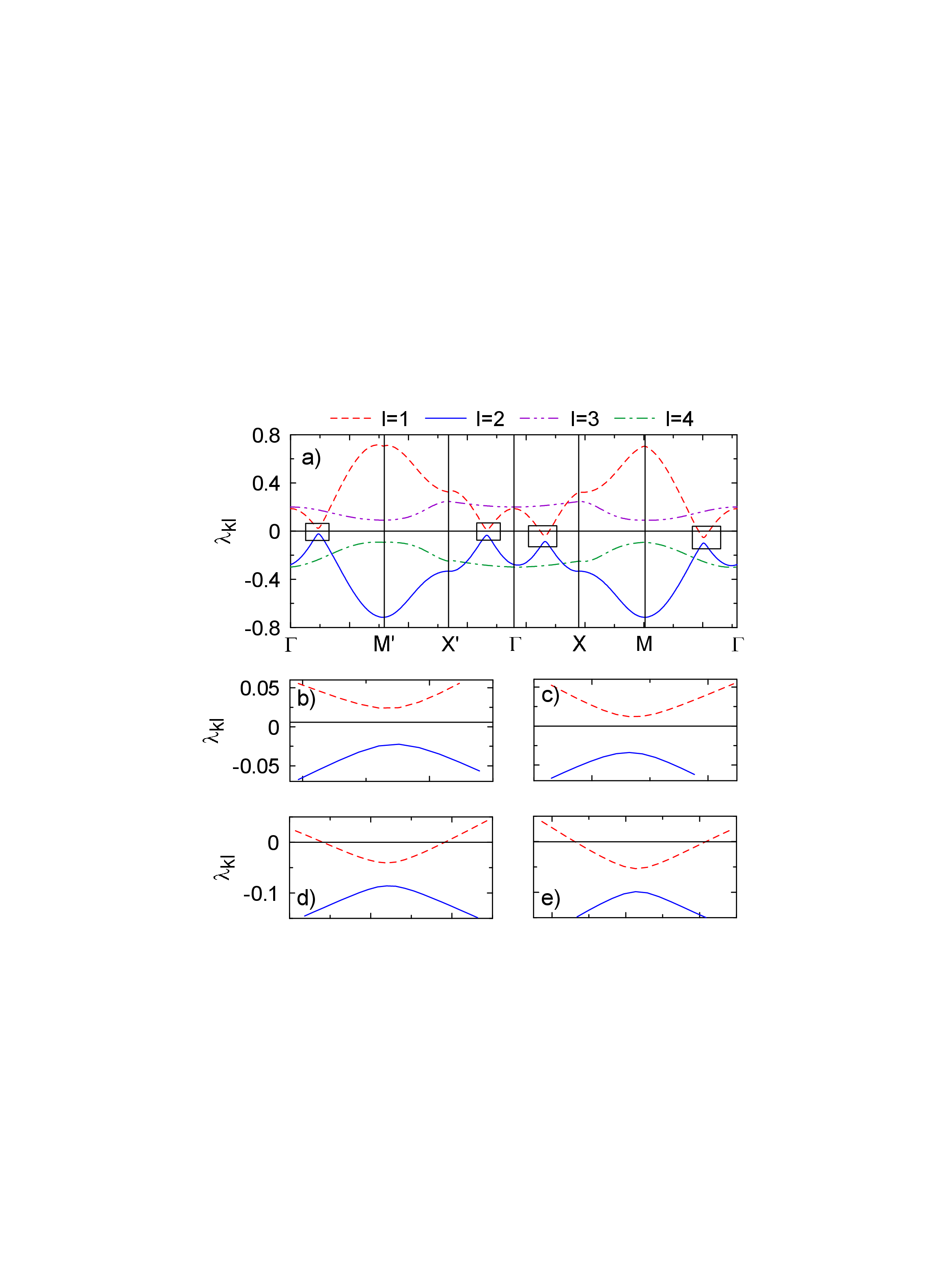}}
\caption{Quasiparticle energies from Eq. (\ref{eq:quasipart_ene}) along the trajectory in the reduced Brillouin zone marked by the dashed line in Fig. \ref{fig:10} for the value of the $Q_x$ component which corresponds to the free energy minimum shown in \ref{fig:8}d. In (b)-(e) the quasiparticle energies are shown zoomed around four points in the Brillouin zone marked by squares in (a) from left to right, respectively. As shown in (d) and (e) the quasiparticle energies intersect the Fermi energy at certain points in the Brillouin zone what leads to the depairing region presented in Fig. \ref{fig:10}.}
\label{fig:9}
\end{figure}

\begin{figure}[h!]
%\hfill
\centering
\epsfxsize=50mm 
{\epsfbox[200 493 434 711]{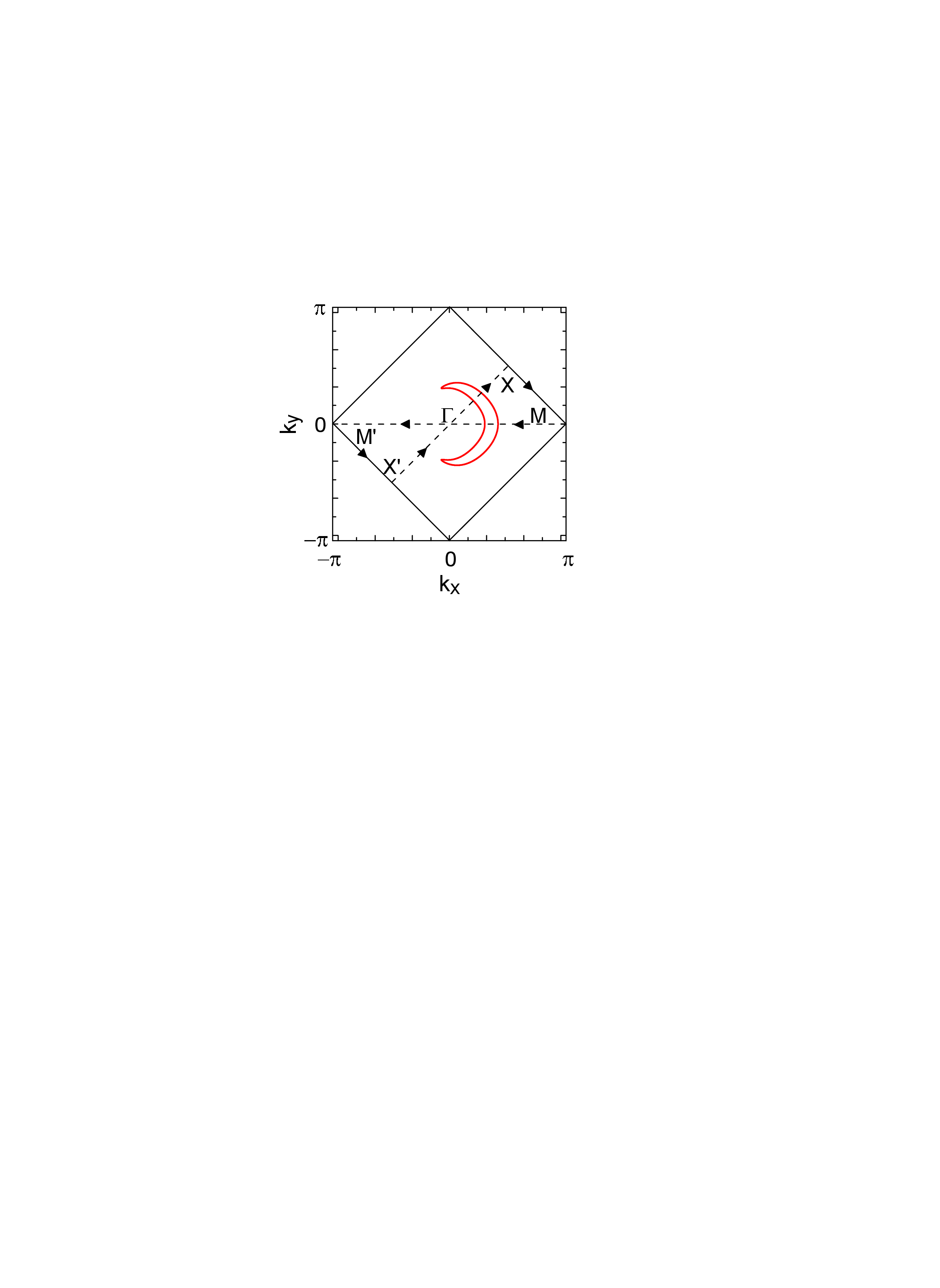}}
\caption{Depairing region in reciprocal space which corresponds to the free energy minimium shown in \ref{fig:8}d.}
\label{fig:10}
\end{figure}

\begin{figure}[h!]
%\hfill
\centering
\epsfxsize=85mm 
{\epsfbox[17 176 598 704]{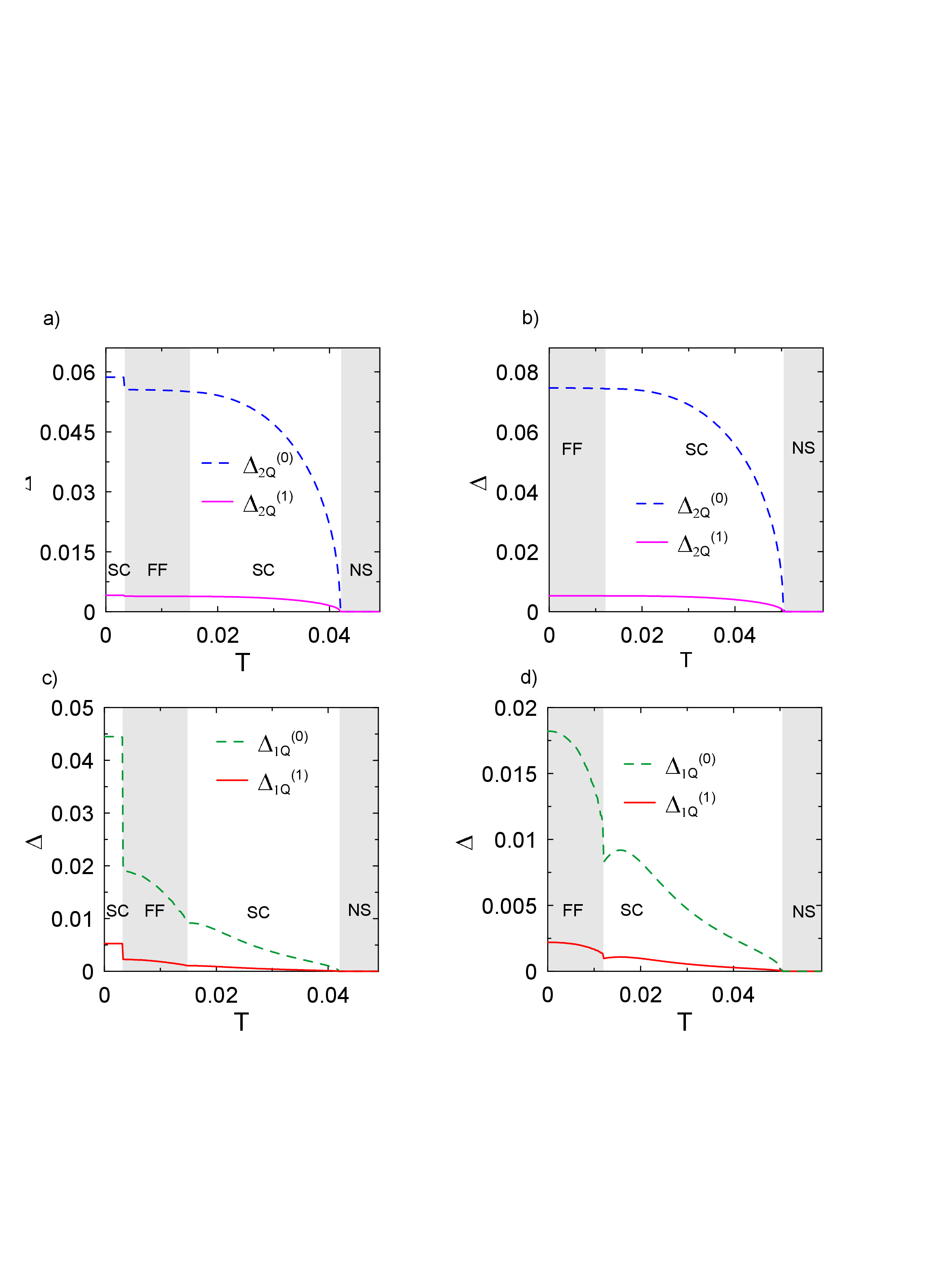}}
\caption{Temperature dependences of the superconducting gaps for two values of the band filling $n=1.72$ (a), (c) and $n=1.795$ (b), (d) for the model with two electron- and two hole-like bands. The pairing strength is set to $U_0=0.33$.}
\label{fig:11}
\end{figure}

In Fig. \ref{fig:8} the free energy in the paired phase as a function of the $\mathbf{Q}$ vector is displayed for the two selected values of the band filling. Analogously to the previous case, the free energy mimima are visible for the proper value of $n$, which signal the FF phase appearance. However, as can be seen from Fig. \ref{fig:9}, even in the FF phase all the particles from the electron-like bands are paired. As a result, the so-called depairing region shown in Fig. \ref{fig:10}, appearance of which is a characteristic feature of the FF phase \cite{Fulde1964,Wu2013}, corresponds to the unpaired particles from the hole-like bands only. This is caused by the fact that the gap amplitudes in the electron-like bands are significantly larger than those corresponding to the hole-like bands (cf. Fig. \ref{fig:11}). Moreover, due to the presence of the intersection points of the electron-like Fermi surface sheets (cf. also Fig. \ref{fig:1}b), the wave vector mismatch between them is highly anisotropic and 
only its maximal value $\Delta k_1$ is close to the mismatch between the 
hole-like Fermi surfaces. In such conditions, the electron-like Fermi surface sheets can still be connected without the necessity of non-zero momentum pairing in the considered parameter range. Hence, the appearance of the FF phase is due to the hole-like bands which have been considered separately in the previous subsection. 
%%%%%%%%%%%%%%%%%%%%%%%%%%%%%%%%%%%%%%%%%%%%FIG9%%%%%%%%%%%%%%%%%%%%%%%%%%%%%%%%
%%%%%%%%%%%%%%%%%%%%%%%%%%%%%%%%%%%%%%%%%%%%%%%%%%%%%%%%%%%%%%%%%%%%%%%%%%%%%%%%
%%%%%%%%%%%%%%%%%%%%%%%%%%%%%%%%%%%%%%%%%%%%FIG10%%%%%%%%%%%%%%%%%%%%%%%%%%%%%%%%
%%%%%%%%%%%%%%%%%%%%%%%%%%%%%%%%%%%%%%%%%%%%%%%%%%%%%%%%%%%%%%%%%%%%%%%%%%%%%%%%

\begin{figure}[h!]
%\hfill
\centering
\epsfxsize=75mm 
{\epsfbox[218 530 462 664]{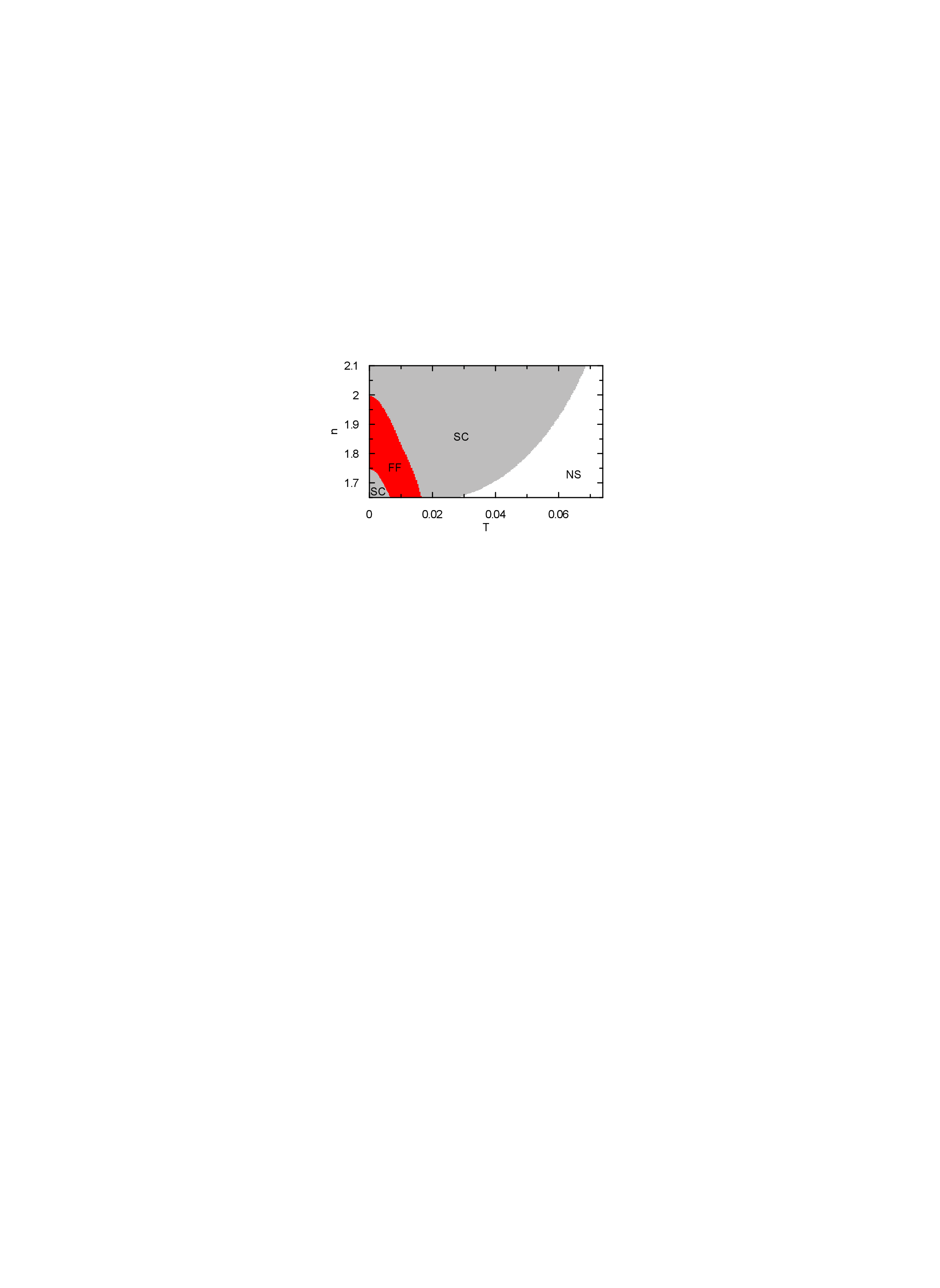}}
\caption{Phase diagram on $(T, n)$ plane for $U_0=0.33$ for the case of the four-band Hamiltonian. Labels of the phases are explained in main text. Note that the FF phase appears only in a narrow range of the band filling.}
\label{fig:12}
\end{figure}

\begin{figure}[h!]
%\hfill
\centering
\epsfxsize=85mm 
{\epsfbox[40 320 608 569]{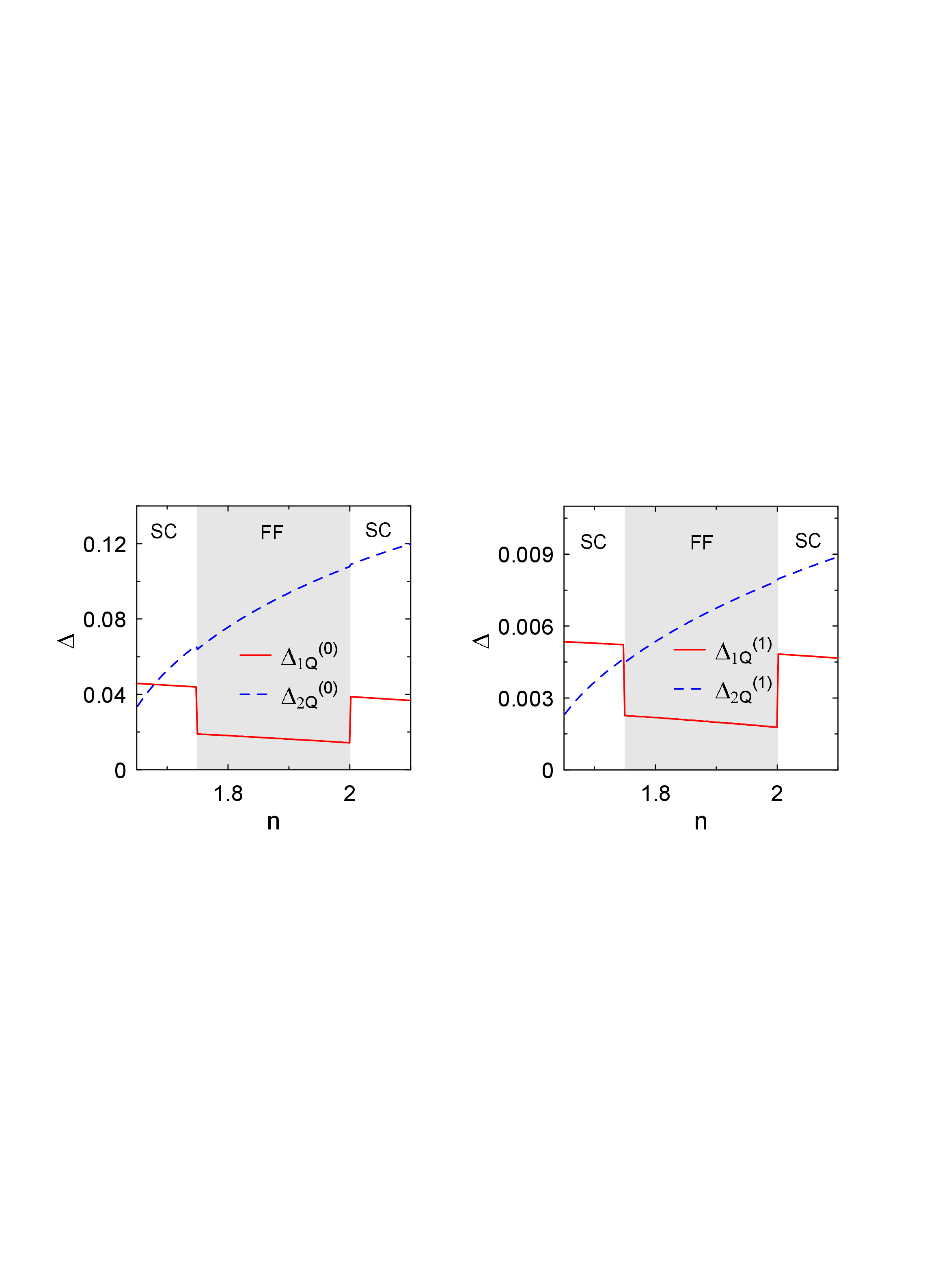}}
\caption{Band filling dependence of the gap components: $s-wave$ (a) and $extended$ $s-wave$ (b) for $U_0=0.33$.}
\label{fig:13}
\end{figure}

The temperature dependences of the superconducting gaps are plotted in Fig. \ref{fig:11}. The transition from SC to FF phase marked in Figs. \ref{fig:11}a and \ref{fig:11}c is of the first order, as in the simplified case of two hole-like bands (cf. Fig. \ref{fig:5}). However, here a transition from the FF to the SC phase can also appear at nonzero temperature. The unconventional temperature dependence of the gaps $\Delta^{(0)}_{1\mathbf{Q}}$ and $\Delta^{(1)}_{1\mathbf{Q}}$ should also be noted. As one can see from the phase diagram in Fig. \ref{fig:12} the stability range of the paired phase is significantly broadened by the presence of the electron-like bands (cf. Fig. \ref{fig:4} for the case of the model with hole-like bands only). However, the dominant part of the diagram in the considered parameter range is covered by the zero-momentum paired phase (SC), whereas the stability region of the FF phase, in spite of some differences, has similar shape and size to that appearing in Fig. \ref{fig:4}. 
This is caused by the fact that the hole-
like bands are responsible for the non-zero momentum pairing. In Fig. \ref{fig:13} one can see that the gap amplitudes corresponding to the electron-like bands increase with increasing $n$ what leads to increasing transition temperature from SC to NS phase seen on the diagram. However, the transition temperature from FF to SC phase decreases with increasing $n$ in a similar manner to the case of the two-band model.

%%%%%%%%%%%%%%%%%%%%%%%%%%%%%%%%%%%%%%%%%%%%FIG11%%%%%%%%%%%%%%%%%%%%%%%%%%%%%%%%
%%%%%%%%%%%%%%%%%%%%%%%%%%%%%%%%%%%%%%%%%%%%%%%%%%%%%%%%%%%%%%%%%%%%%%%%%%%%%%%%
%%%%%%%%%%%%%%%%%%%%%%%%%%%%%%%%%%%%%%%%%%%%FIG12%%%%%%%%%%%%%%%%%%%%%%%%%%%%%%%%
%%%%%%%%%%%%%%%%%%%%%%%%%%%%%%%%%%%%%%%%%%%%%%%%%%%%%%%%%%%%%%%%%%%%%%%%%%%%%%%%

%%%%%%%%%%%%%%%%%%%%%%%%%%%%%%%%%%%%%%%%%%%%FIG13%%%%%%%%%%%%%%%%%%%%%%%%%%%%%%%%
%%%%%%%%%%%%%%%%%%%%%%%%%%%%%%%%%%%%%%%%%%%%%%%%%%%%%%%%%%%%%%%%%%%%%%%%%%%%%%%%

\section{Conclusions and outlook}
We propose a new kind of nonzero-momentum paired phase which can appear spontaneously (without any external magnetic field). This kind of phase can be created in the systems with dominant interband pairing for which a sufficient Fermi vector mismatch appears and can be compensated by a nonzero momentum of the Cooper pairs. It should be noted that in such scenario, the requirement of having a high value of the Maki parameter, to make the observation of conventional FFLO phase possible, does not appear at all. We analyze the proposed idea for the case of a relatively simple model which is based on two hole-like and two electron-like bands appropriate for the iron based superconductor LaFeAsO$_{1-x}$F$_x$. We show that by doping the system with electrons one can tune the distance between the Fermi surface sheets and, in result, create favorable conditions for the appearance of this unconventional phase. For appropriate values of the band filling, the free energy minima appear which 
signal the possibility of this inhomogeneous paired state (cf. Figs. \ref{fig:3}, \ref{fig:8}). The FF phase is stable in a fairly narrow band-filling range (cf. Figs. \ref{fig:4} and \ref{fig:12}), which may pose 
stringent conditions on its detection. 

The temperature dependence of the superconducting gaps in the limited case of two hole-like bands only, is similar to that for the case of standard magnetic-field-induced FFLO state, with first-order transition from a homogeneous to an inhomogeneous paired phase (cf. Fig. \ref{fig:5}). In the extended four-band model some unusual behavior of the gap amplitudes as functions of the temperature should be noted, as well as the appearance of the the FF to SC phase transition. Moreover, for the case of the full four-band Hamiltonian the stability range of the superconducting phases is broadened mainly by the zero-momentum paired phase (see Fig. \ref{fig:12}) which is due to the presence of the electron-like bands with high value of the gap amplitudes relative to those coming from the hole-like bands. The shape and size of the area corresponding to the stability of the Fulde-Ferrell phase is similar in both considered situations (the two-band and the four-band cases as presented in Figs. \ref{fig:4} and \ref{fig:12}
, respectively). This is caused by the fact that even in the case of the four-band model the particles from the hole-like bands are responsible for the non-zero momentum paired state appearance. This is because the Fermi wave vector mismatch between the electron-like bands is much smaller (or even vanishes) than that between the hole-like bands for large parts of the Fermi surfaces (cf. Fig. \ref{fig:1}b). This together with relatively high values of the gap amplitudes $\Delta^{(0)}_{2\mathbf{Q}}$ and $\Delta^{(0)}_{2\mathbf{Q}}$, leads to the situation in which the non-zero momentum pairing  is unlikely to be induced by the mismatch between the electron-like Fermi surfaces. 

In a similar model but for conditions in which the FFLO state can be triggered due to both the electron- and the hole-like bands, but with different favorable $\mathbf{Q}$ vectors, a competition between two non-zero momentum paired phases would appear (each corresponding to different value of the modulation). Such competition is considered in a two band model in Ref. \onlinecite{Takahashi2014} but for the case of standard magnetic effect induced FFLO phase. 

The presented mechanism which leads to a spontaneous appearance of the non-zero momentum paired phase can be studied for varius systems with an interband pairing. In this respect the new class of ``Hund's metals'' \cite{Hardy2013} should be mentioned. The Hunds rule induced spin-triplet superconducting phase with an interband pairing has been discussed by us recently in the correlated regime\cite{Spalek2013,Zegrodnik2014}, though without taking into account the possibility of non-zero momentum pairing. In this analysis only two equivalent bands with hybridization between them have been considered. Due to the hybridization term the bonding and antibonding bands can be created what leads to a Fermi wave vector mismatch similar as in the case of the two hole-like bands studied here (Fig. 3b from Ref. \onlinecite{Spalek2013}). It should be possible to obtain the stability of a spontaneous FFLO phase also in that model. Moreover, systems with pairing between two types of particles with different effective 
masses, even in the case of parabolic bands, can also be analyzed with respect to the inhomogeneous superconducting phase appearance in the absence of magnetic field. The different shapes of the dispersion relations which correspond to the two species of particles can lead to a Fermi wave vector mismatch which in turn can induce non-zero momentum pairing of either spin-singlet of spin-triplet type. 

The \textit{sine qua non} condition for the appearance of the proposed phase is the dominant role of the interband pairing. In this respect, the phonon coupling cannot differentiate essentially between the intra- and the inter-band couplings, whereas the magnetic Hund's rule coupling does favor the interband interaction correlations. To expose the role of the interband pairing mechanism, here we have assumed that only the particles from two different Fermi surface sheets can be paired. Nevertheless, one can expect that in the case with both inter- and intra-band pairing, the energy gain coming from the non-zero 
momentum pairing between the bands would not be completely reduced by the corresponding contributions from intraband coupling. On the other hand, when the latter is strong enough, the homogeneous superconducting state is to be favored. This last issue should be analyzed separately.

\section{Acknowledgement}
Discussions with Jan Kaczmarczyk and Andrzej Ptok are gratefully acknowledged. The authors are grateful to the Foundation for Polish Science (FNP) for support within the project TEAM, as well as to the National Science Center (NCN) through Grant MAESTRO, No. DEC-2012/04/A/ST3/00342.


\begin{thebibliography}{99}

\bibitem{Fulde1964}
P. Fulde and R. A. Ferrell, Phys. Rev. {\bf 135}, A550 (1964).

\bibitem{Larkin1964}
A. I. Larkin and Y. N. Ovchinnikov, Sov. Phys. JETP {\bf 20}, 762 (1964).

\bibitem{SaintJames1969}
D. Saint-James, G. Sarma, and E. J. Thomas: \textit{Type II Superconductivity} (Pergamon, New York, 1969).

\bibitem{Gloos1993} %heavy fermion FFLO experimental
K. Gloos, R. Modler, H. Schimanski, C. D. Bredl, C. Geibel, F. Steglich, A. I. Buzdin, N. Sato, and T. Komatsubara, Phys. Rev. Lett. {\bf 70}, 501 (1993).

\bibitem{Bianchi2003} %heavy fermion FFLO experimental
A. Blanchi, R. Movshovich, C. Capan, P. G. Paglluso, and J. L. Sarrao, Phys. Rev. Lett. {\bf 91}, 187004  (2003).

\bibitem{Kumagai2006} %heavy fermion FFLO experimental
K. Kumagai, M. Saitoh, T. Oyaizu, Y. Furukawa, S. Takashima, M. Nohara, H. Takagi, Y. Matsuda, Phys. Rev. Lett. {\bf 97}, 227002 (2006).

\bibitem{Correa2007} %heavy fermion FFLO experimental
V. F. Correa, T. P. Murphy, C. Martin, K. M. Purcell, E. C. Palm, G. M. Schmiedenshoff, J. C. Cooley, and W. Tozer, Phys. Rev. Lett. {\bf 98}, 087001 (2007).

\bibitem{Lee1997} %organic FFLO experimental
I. J. Lee, M. J. Naughton, G. M. Danner, and P. M. Chaikin, Phys. Rev. Lett. {\bf 78}, 3555 (1997).

\bibitem{Singleton2000} %organic FFLO experimental
J. Singleton, J. A. Symington, M.-S. Nam, A. Ardavan, M. Kurmoo, and P. Day, J. Phys.: Condens. Matter {\bf 12}, L641 (2000).

\bibitem{Tanatar2002} %organic FFLO experimental
M. A. Tanatar, T. Ishiguro, H. Tanaka, and H. Kobayashi, Phys. Rev. B {\bf 66}, 134503 (2002).

\bibitem{Uji2006} %organic FFLO experimental
S. Uji, T. Terashima, M. Nishimura, Y. Takahide, T. Konoike, K. Enomoto, H. Cui, H. Kobayashi, A. Kobayashi, H. Tanaka, M. Takumoto, E. S. Choi, T. Tokumoto, D. Graf, and J. S. Brooks, Phys. Rev. Lett. {\bf 97}, 157001 (2006).

\bibitem{Shinagawa2007} %organic FFLO experimental
J. Shinagawa, Y. Kurosaki, F. Zhang, C. Parker, S. E. Brown, D. J\'erome, J. B. Christensen, and K. Bechgaard, Phys. Rev. Lett. {\bf 98}, 147002 (2007).

\bibitem{Ptok2013_1}
A. Ptok and D. Crivelli, J. Low Temp. Phys. {\bf 172}, 226 (2013).

\bibitem{Ptok2014_1}
A. Ptok, Eur. Phys. J. B {\bf 87}, 2 (2014).

\bibitem{Ptok2014_2}
D. Crivelli and A. Ptok, arxiv:1401.3066v1 (unpublished).

\bibitem{Zocco2013}
D. A. Zocco, K. Grube, F. Eilers, T. Wolf, and H. v. L{\"o}hneysen, Phys. Rev. Lett {\bf 111}, 057007 (2013).

\bibitem{Zwierlein2006} %imbalanced ultracold Fermi gases
M. W. Zwierlein, A. Schirotzek, C. H. Schunk, W. Ketterle, Science {\bf 311}, 492 (2006).

\bibitem{Schunck2007} %imbalanced ultracold Fermi gases
C. H. Schunck, Y. Shin, A. Schirotzek, M. W. Zwierlein, and W. Ketterle, Science {\bf 316}, 867 (2007).

\bibitem{Liao2010}  % FFLO in the 1D ultracold Fermi gas
Y. Liao, A. S. C. Rittner, T. Paprotta, W. Li, G. B. Partridge, R. G. Hulet, S. K. Baur, and E. J. Mueller, Nature {\bf 467}, 567 (2010).

\bibitem{Wu2013}  %FFLO induced by artificial spin-orbit coupling
F. Wu, G.-C.Guo, W. Zhang, W. Yi, Phys. Rev. Lett. {\bf 110}, 110401 (2013).

\bibitem{Liu2013}  %FFLO induced by artificial spin-orbit coupling
X.-J. Liu and H. Hu, Phys. Rev. B {\bf 87}, 051608(R) (2013).

\bibitem{Dong2013}  %FFLO induced by artificial spin-orbit coupling
L. Dong, L. Jiang, and H. Pu, New J. Phys. {\bf 15}, 075014 (2013).

\bibitem{Tanaka2007}
H. Tanaka, H. Kaneyasu, and Y. Hasegawa, J. Phys. Soc. Jpn. {\bf 76}, 024715 (2007).

\bibitem{Loder2010}
F. Loder, A. P. Kampf, and T. Kopp, Phys. Rev. B {\bf 81}, 020511(R) (2010).

\bibitem{Nikolic2010}
P. Nikoli\'c , A. A. Burkov, and A. Paramekanti, Phys. Rev. B {\bf 81} 012504 (2010).

\bibitem{Mazin2008} %spin-singlet LaFeAsO
I.I. Mazin, D. J. Singh, M. D. Johannes, and M. H. Du, Phys. Rev. Lett. {\bf 101}, 57003 (2008).

\bibitem{Kuroki2008} %spin-singlet LaFeAsO (s-wave, d-wave)
K. Kuroki, S. Onari, R. Arita, H. Usui, Y. Tanaka, H. Kontani, and H. Aoki, Phys. Rev. Lett. {\bf 101}, 087004 (2008).

\bibitem{Chen2009} %spin-singlet LaFeAsO
W.-Q. Chen, K.-Y. Zhou, and F.-C. Zhang, Phys. Rev. Lett. {\bf 102}, 047006 (2009).

\bibitem{Maier2008} %spin-triplet and spin-singlet LaFeAsO
T. A. Maier and D. J. Scalpino, Phys. Rev. B {\bf 78}m 020514(R) (2008).

\bibitem{Lee2008} %spin-triplet LaFeAsO
P. A. Lee and X.-G. Wen, Phys. Rev. B {\bf 78}, 144517 (2008).

\bibitem{Dai2008}  %spin-triplet LaFeAsO
X. Dai, Z. Fang, Y. Zhou, and F.-C. Zhang, Phys. Rev. Lett. {\bf 101}, 057008 (2008).

\bibitem{Yashima2009}
M. Yashima, H. Nishimura, H. Mukuda, Y. Kitaoka, K. Miyazawa, P. M. Shirage, P. M. Shirage K. Kihou, H. Kito, H. Eisaki, and A. Iyo, J. Phys. Soc. Jpn. {\bf 78}, 103802 (2009).

\bibitem{Jeglic2010}
P. Jegli\v c, A. Poto\v cnik, M. Klanj\v sek, M. Bobnar, M. Jagodi\v c, K. Koch, H. Rosner, S. Margadonna, B. Lv, A. M. Guloy, and D. Ar\v con, Phys. Rev. B. {\bf 81}, 140511(R) (2010).

\bibitem{Michioka2010}
C. Michioka, H. Ohta, M. Matsui, J. Yang, K. Yoshimura, and M. Fang, Phys. Rev. B {\bf 82}, 064506 (2010).

\bibitem{Singh2008} %band structure calculations LaFeAsO
D. J. Singh and M.-H. Du, Phys. Rev. Lett. {\bf 100}, 237003 (2008).

\bibitem{Xu2008} %band structure calculations LaFeAsO
G. Xu, W. Ming, Y. Yao, X. Dai, S.-C. Zhang, and Z. Fang, Europhys. Lett. {\bf 82}, 67002 (2008).

\bibitem{Raghu2008} %minimal two-band model for LaFeAsO
S. Raghu, X.-L. Qi, C.-X. Liu, D. J. Scalpino, and S.-C. Zhang, Phys. Rev. B {\bf 77}, 220503(R) (2008).

\bibitem{Ran2009}  %two-band model fo LaFeAsO
Y. Ran, F. Wang, H. Zhai, A. Visghwanath, and D.-H. Lee, Phys. Rev. B {\bf 79}, 014505 (2009).

\bibitem{Spalek2001}
J. Spa{\l}ek, Phys. Rev. B {\bf 63}, 104513 (2001).

\bibitem{Zegrodnik2012}
M. Zegrodnik, J. Spa{\l}ek, Phys. Rev. B {\bf 86}, 014505 (2012).

\bibitem{Zegrodnik2013}
M. Zegrodnik, J. Spa\l ek, and J. B\"unemann, New J. Phys. {\bf 15}, 07305 (2013).

\bibitem{Spalek2013}
J. Spa\l ek, M. Zegrodnik, J. Phys.: Condens. Matter {\bf 25}, 435601 (2013).

\bibitem{Zegrodnik2014}
M. Zegrodnik, J. B\"unemann, and J. Spa\l ek, New J. Phys {\bf 16}, 033001 (2014).

\bibitem{Kaczmarczyk2010}
J. Kaczmarczyk and J. Spa\l ek, J. Phys. Condens. Matter {\bf 22} 355702 (2010).

\bibitem{Maska2010}      % wykres Q(T,H)
M. M. Ma{\'s}ka, M. Mierzejewski, J. Kaczmarczyk, J. Spa{\l}ek, Phys. Rev. B {\bf 82}, 054509 (2010).

\bibitem{Hardy2013}
F. Hardy, A. E. B{\"o}hmer, D. Aoki, P. Burger, T. Wolf, P. Schweiss, R. Heid, P. Adelmann, Y. X. Yao, G. Kotliar, J. Schmalian, and C. Meingast, Phys. Rev. Lett. {\bf 111}, 027002 (2013).

\bibitem{Takahashi2014}
M. Takahashi, T. Mizushima, and K. Machida, Phys. Rev. B {\bf 89}, 064505 (2014).




%\bibitem{Shimahara1994}  %FFLO - phase diagram -Fig. 2
%H. Shimahara, Phys. Rev. B {\bf 50}, 12760 (1994).



%\bibitem{Dupuis1995} %praca teoretyczna o quasi 1D FFLO paired phase, odnosza to do organic supercond.
%N. Dupuis, Phys. Rev. B {\bf 51}, 9074 (1995).

%\bibitem{Buzdin1997}
%A. I. Buzdin and H. Kachkachi, Phys. Lett. A {\bf 225}, 341 (1997).




\end{thebibliography}
\end{document}